# An Artificial Immune based Approach for Detection and Isolation Misbehavior Attacks in Wireless Networks

Shahram Behzad and Reza Fotohi[1*], jaber hosseini balov[2], Mohammad javad Rabipour[3]

[1] Department of Computer Engineering Germi branch, Islamic Azad University, Germi, Iran
[2] Department of computer Engineering, Information Technology and Management, Jönköping University, Sweden
[3] Department of Electrical & Computer Engineering, Kashan Branch, Islamic Azad University, Kashan, Iran

**Abstract:** MANETs (Mobile Ad-hoc Networks) is a temporal network, which is managed by autonomous nodes, which have the ability to communicate with each other without having fixed network infrastructure or any central base station. Due to some reasons such as dynamic changes of the network topology, trusting the nodes to each other, lack of fixed substructure for the analysis of nodes' behaviors and loss of specific offensive lines, this type of networks is not supportive against malicious nodes' attacks. One of these attacks is black hole attack. In this attack, the malicious nodes absorb data packets and destroy them. Thus, it is essential to present an algorithm against the black hole attacks. This paper proposed a new approach, which improvement the security of DSR routing protocol to encounter the black hole attacks. This schema tries to identify malicious nodes according to nodes' behaviors in a MANETs and isolate them from routing. The proposed protocol, called AIS-DSR (Artificial Immune System DSR) employ AIS (Artificial Immune System) to defend against black hole attacks. AIS-DSR is evaluated through extensive simulations in the ns-2 environment. The results show that AIS-DSR outperforms other existing solutions in terms of throughput, end-to-end delay, packets loss ratio and packets drop ratio.

## 1. Introduction

Mobile Ad Hoc networks do not have any access point to network accessibility. Wireless devices such as notebooks, laptop and cell phones connect to similar equipment and form an Ad Hoc network. Since nodes are not controlled by any central entity, they have unrestricted mobility and connectivity to others. Routing and network management are done by each other nodes, cooperatively. In other words, the nodes' communications are formed based on the cooperation and some trust among them. In these networks, each node works as a host as well as a router that forwards packets for other nodes. The most important property of these networks is their dynamic and variable topologies that is the result of the nodes' mobility [1, 2 and 3]. Security is one of the main research topics in computer networks. The wide usage of mobile ad hoc networks in martial environment and other security sensitive usages have made the security a basic requirement for these networks. Because nodes participate in the routing process, they can destroy the network. As routing is based on some kind of trust between nodes, it provides a good chance for attackers to disorder routing process. As these networks are formed without any pre plan and for a short period; therefore, the security in these networks is investigated separately. One of the most famous attacks is Black Hole attack. A black hole is a malicious node that replies for any route requests without having any active route to specified destinations and purges all the receiving packets. In addition, in black Hole, attacker nodes tend to advertise and spread fake routes, absorb network traffic towards their selves and drop packets. Fig. 1. Shows an example of black hole attack.

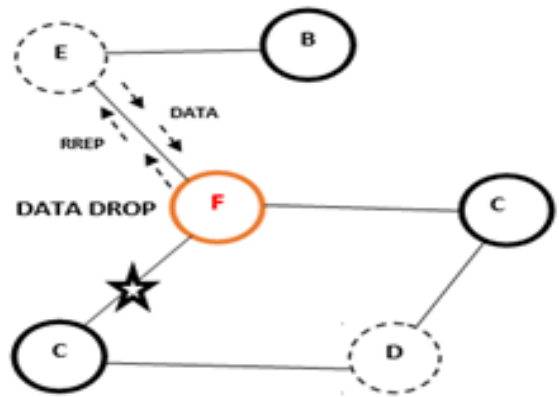

Fig. 1. MANETs with wormhole attack.

basic concepts and preliminaries including DSR routing protocol, black hole attack and artificial immune system are provided in Sects. 2, 3 and 4, respectively. Section 5 discusses the related works. The proposed approach is given in Sect. 6. Performance evaluation are presented in Sect. 7 and finally Sect. 8 concludes the paper.

## 2. DSR ROUTING PROTOCOL

DSR routing protocol follows the basic, When RREQ process is sent, the node waits for RREP and once the RREPs come from the nodes, it responses to the first arrived RREP (Fig. 2). The node sends packets with this RREP, which in turn leads to ignoring other REEPs. Such a process leads to ignoring the security or in security of the route, which in turn enhance the chance of malicious nodes (black hole nodes) existence eliminating the transmitted packets. To avoid these types of attacks in the network, a method based on table and count hope, imitating the artificial immune system, is used. Through this process, we demand for a route and wait for the node response to transmit the packets. Once a large number of RREPs are received in a successive time period from a node, the packets are not transmitted with this RREP. Rather, information of the given node is recorded in a table and, to evaluate the immunity of the route, the hop counts transmitted by RREP are analyzed. In the case RREP hop counts are less, for instance there exists a node with high RREPs and also less hop counts, the given node is labeled as suspicious in the table and its information is broadcasted to the neighboring nodes , that node is eliminated from operation cycle (Fig. 3).

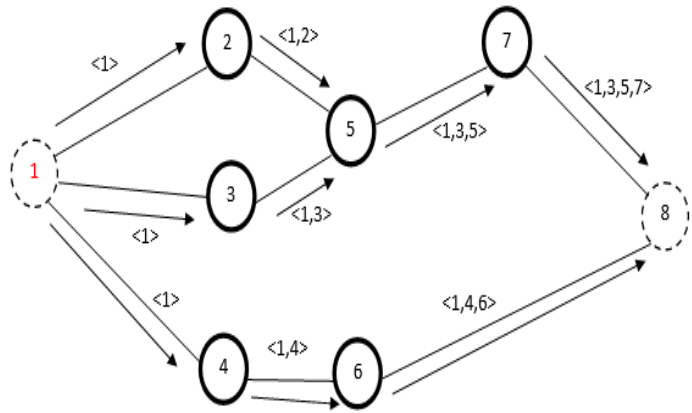

Fig. 2. DSR Route Request (RREQ)

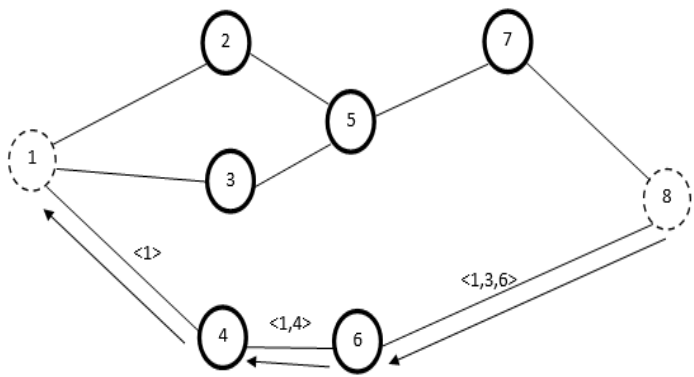

Fig. 3. DSR Route Reply (RREP)

The black hole attack should enter to broadcast group to be able to separate the packs from the multicast. This kind of attack deletes some or the entire recipient packs instead of sending them and consequently it makes the result of packet delivery rate low. Black hole attack is divided into two groups: single black hole attack is applied via one of the existence nodes in the network and total black hole attack that corporates together for an attack more than one node [4]. The node, which executes the black hole attack, waits to receive a RREQ. With the receipt of this RREQ, the invader node without regarding its routing list or even considering any path to the destination, answers positively to that recipient RREQ and makes sending this RREP shorter than other nodes. The invader node cheats the node, which sends the RREQ by putting the most number of sequences and the least number of hops in the RREP. The node which sends the RREQ supposes that it has found the best path when receiving this RREP. Therefore, this node is considered as a short and fit path to send the packet. Because of that a black hole is created and each node- which is known as a black hole instead of sending the packets to destination gets their data or throws them away. The attack node does not check its routing list, so it answers to the RREQ node before the other nodes. If the attack node introduces itself as a fit path for the total network nodes and succeed to gain all network traffic, so it can destroy all network paths and prepare a DOS attack [5] (Fig. 4).

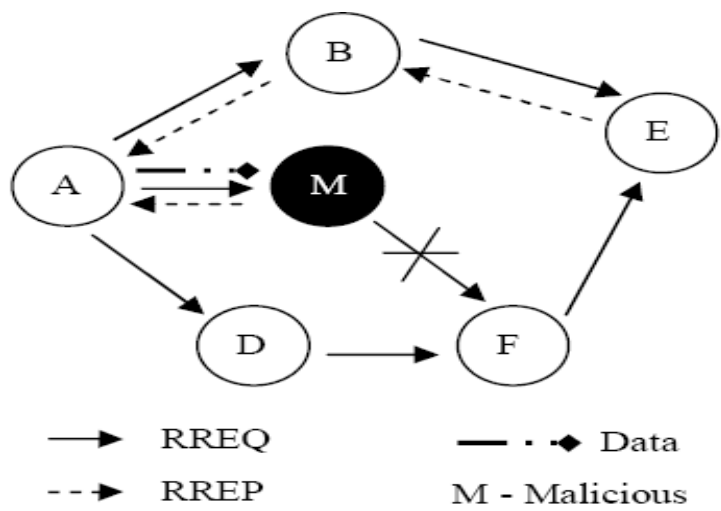

Fig. 4. DSR Route Reply (RREP)

## 3. THE ARTIFICIAL IMMUNE SYSTEM

In general, the artificial immune systems (AIS) [1 and 6] are categorized on inspired algorithms from the biology. As the name would imply, these kinds of algorithms are computer based algorithms which their features and principles are the result of close examination in both adaptive characteristics 3 and the resistance of biological samples4. Samples of such algorithms are brain-inspired neurotic networks, as well as artificial immune system, which make use of principles and natural immune system processes to solve problems.

### 3.1. Immunity

In this section, the immune principles would be insinuated in such a way that will facilitate understanding the rest of paper and we avoid giving excess details. This section is beginning with a short summery about intrinsic immune system. The intrinsic immune system has not been used extensively in artificial systems, however, due to assisting the performance of acquisitive immune system and affecting it and the usage of it in this paper, having a short review regarding that seems necessary.

### 3.2. Intrinsic immunity

As the name would suggest, the intrinsic immune system would last not change over time and been adjusted for detecting a small number of common aggressors. The intrinsic immune system will diminish most of pathogens (which are the potential detrimental aggressors) at its first clash. The acquisitive immune system is in desperate needs of time to response against aggressors, so it is the intrinsic immune system responsibility to react against the aggressors immediately and bring the attack under its control   so as to the acquisitive immune system could give an influential response [1].

### 3.3.    The acqustive immunity

When the intrinsic immunity is made active, its activity would last for several days while the time the acquisitive immunity is active, it would even endure for weeks. The acquisitive immunity is obliged to destroy the pathogens in case the intrinsic immunity has been defeated or is infeasible anyway. Inapplicability of the intrinsic immunity is due to its incapability to build a specific response for an aggressor pathogen; in this point, it is the acquisitive immune system, which takes the field. Unlike the intrinsic immunity, the acquisitive immunity is specific and has memory, according to Fig. 5, it can remember the pathogen which has attacked one time and system generated a response for it, therefore in future encounters it can respond quickly to oppose it.

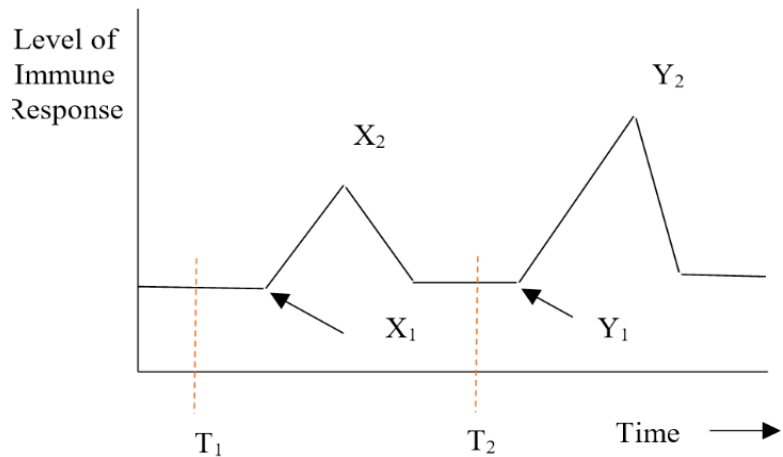

Fig. 5. DSR Route Reply (RREP)

Response of $Y_2$ is stronger than $X_2$. The first sign for existence of such unknown antigens in $T_1$ produces $X_2$ response. Nevertheless, notice the delay between $T_1$ & $X_1$. Nearly The same antigens been interred in the time of $T_2$, almost immediately in produced response of $Y_1$ for acquisitive system can be observed in vaccination experiments carried out by Jenner in 1970 [1 and 8].

### 3.4.    B-cell, antibody and immunoglobulin

Like all immune cells, B cells are developed in bone marrow. A natural B cell contains 105 antibodies (receptor) on its surface. Each one of these antibodies has a unique shape that can be found in genetic structure of B cell and as a result they have a shape similar to B cell. Therefore all antibodies produced through a B cell are connected to a similar set of molecular models. According to figure 4, antibodies of B cell are dual-value and dual- performance. They are dual-value because of their capability for connecting to two antigens through two arms in Fab areas, and they are called dual- performance due to the fact that besides having the capability for connecting to antigen models using Fab areas, they can also be connected to certain receptors on surface of other immune cells using Fc portion [2 and 7].

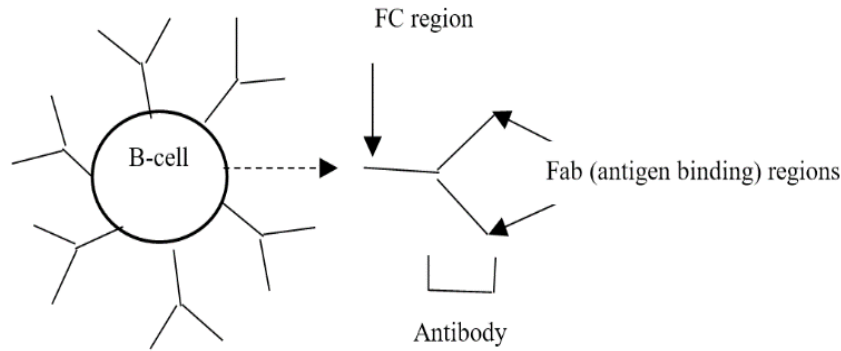

Fig. 6. B-cell, antibody and immunoglobulin

There are copious antibodies on the surface of B-cell. In Fig. 6, an antibody (Ab), or an immunoglobulin (Ig), is consisted of two similar light chains (L) and two similar heavy chains (H).

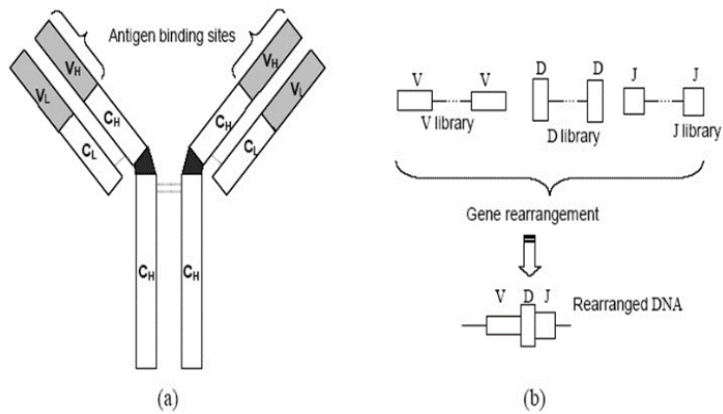

Fig. 7. Antibody molecule and its genome

## 4. RELATED WORKS

Computer networks are threatened by many security attacks such as modification, denial of service attack, fabrication attack, IP spoofing, etc. Therefore, a lot of research has been done in this regard [9–15].

Rutvij et al. [16] proposed a technique which makes use of AODV protocol to detect Black Hole attack. In this work, a dynamic node calculates a peak value for every time interval. The peak value calculates RREP sequence number, routing table sequence number and number of replies received during a particular time interval. These values are calculated for every time interval and in this way the attacks are calculated dynamically. According to their algorithm when an intermediate node receives RREP packets having higher sequence number than the peak value that node are marked as malicious node. Further the other nodes in the routing path get the details about malicious node information and isolate that vulnerable node from the network.

Su [17] proposed an intrusion detection technique called an Anti-Black Hole Mechanism (ABM) which employs two additional tables known as RREQ (RQ) table and Sniff Node (SN) table. RQ table maintains all the values of RREQ messages within the transmission range of the node, whereas SN table records the suspicious nodes within its transmission range. The value of the suspicious node is calculated by identifying the forwarded message of intermediate node.

A methodology for identifying multiple black hole nodes cooperating as a group with slightly modified AODV protocol by introducing Data Routing Information (DRI) Table and Cross Checking was proposed [17]. Simulation of this method was implemented in [18].

Authors added some changes to the algorithm to enhance the accuracy in preventing the black hole attacks. They suggested checking the current intermediate node for black hole if the next hop was not reliable. Intrusion detection in sensor networks was studied and a lightweight distributed scheme was proposed [19].

Su [20] extended several intrusion detection system nodes in MANETs in order to detect and prevent selective black hole attacks. A selective black hole is a node which may optionally and alternately do a black hole attack or act as a normal node. The IDS nodes are set in sniff mode in order to estimate the suspicious values of nodes according to the abnormal difference between the routing messages transmitted from the nodes. In case of exceeding threshold, a near IDS asks all nodes on the network to cooperatively isolate the malicious node. The value of the threshold plays an important role in this study. In the proposed method, nodes monitor their neighbourhood and collaborate with their nearest neighbours to return the network back to its normal operational condition. The authors applied their scheme for the black hole and selective forwarding attacks.

In [21], to overcome the black hole attack, authors proposed to ignore the first RREP packet reaching the source node. They implemented this solution by implementing RREP packet caching mechanism. A cluster-based scheme to prevent black hole attacks in MANETs was presented in [22], which elects cluster heads to prevent black hole attacks. Then the authors proposed a cluster-based countermeasure to prevent the black hole attacks by identifying those black hole nodes. In another clustering black hole attack detection and prevention approach, each member of a cluster ping the cluster head to detect the peculiar difference between the numbers of data packets received and forwarded by the node. If anomalousness is detected, all the nodes will delete the malicious nodes from the network [23].

Araghi et al. [24] provided a solution to prevent the black hole attack. In this study the basic authentication of intermediate nodes that send the path response message but get the confirmation from the destination was reviewed. If the confirmation is not received from the destination, these malicious intermediate nodes will be saved for arbitration at a later time. CL parameter is a counter that shows the miss behaviour of intermediate nodes when they send a wrong path response. If CL becomes more than 3 for each node, that node is considred as malicious and the path introduced by this node is avoided.

In [25], an efficient approach for detection and removal of single and cooperative black hole attacks in MANET was presented. The algorithm not only detects the black hole nodes in case when the node is not idle but it can also detect the black hole nodes in case when a node is idle as well. To check if a node is idle, the authors used a threshold for the Interval. In case of exceeding the threshold, the malicious procedure is invoked. For this, a source node looks at its route cache, then sends RREQ packets, and waits for the replies. Based on the replies, the black hole detection procedure is called. In this method, RREQ packets are sent in Fibonacci series pattern until the Flow count Threshold (which is set to 34 ms) is not reached.

Madadian et al. [26] proposed a method to detect black hole attacks. In this method, when a node sends the path response package, a consultation process takes place around that node. Then, according to the comments by neighbor nodes, a decision is made about the maliciousness of the responsive node. One method to detect the black hole attack using a timer was presented.

Mandala [27] proposed new black hole attacks, namely Independent Hybrid Black Hole Attack (IHBHA) and Cooperative Hybrid Black Hole Attack (CHBHA); and then measures the severity of both CBA and HBHAs. Considering the principles and characteristics of the active black hole attacks, an effective approach which can detect and defend active black hole attacks was proposed by improving the AODV routing protocol combining flow analysis [28].

Another method called CDSM (Code Division SecurityMethod) based on code division to avoid the black hole attack was proposed [29] as well. Moreover, some methods to avoid the black hole attacks were compared in [30]. Filling gaps in measuring the severity of this attack,

Another approach was proposed to detect black hole nodes in the MANET. In the proposed method, as soon as detecting a misbehaving node, the detecting node tries to avoid the misbehaving node [31].

## 5. THE PROPOSED APPROACH (AIS-DSR)

In this section, we design a black hole-immune routing protocol by jointly employing the artificial immune system.

### 5.1. Proposed AIS-based Defense scheme against black hole attack

In order to make computer networks intelligent, we can make use of existing knowledge in the nature. One of these systems is the artificial immune system inspired by the human's immune system mechanism, which is an effective solution for convoluted predicaments in computer networks, such as in misbehavior detection or various attacks. In DSR routing protocol, once the first RREP is received by the source node, the packets will be transmitted to the destination node. This shortest route path approach, disregards route's security. In this paper, we do not send immediately packets toward the destination node, upon the first RREP reception. Rather, source node considers entire received PREPs and then selects the most secure route by using the artificial immune system. General concept of this process is like a human being immune system, where antibodies are trained to detect and eliminate malicious antigens. To this end, the proposed approach develops and updates a set of conditions that could detect the routes infected by black hole attackers and abstain from selecting them. As shown in Fig. 8, the proposed scheme has different steps. In the steps, as discussed in section Step 1, the AIS is used to select immune routes.

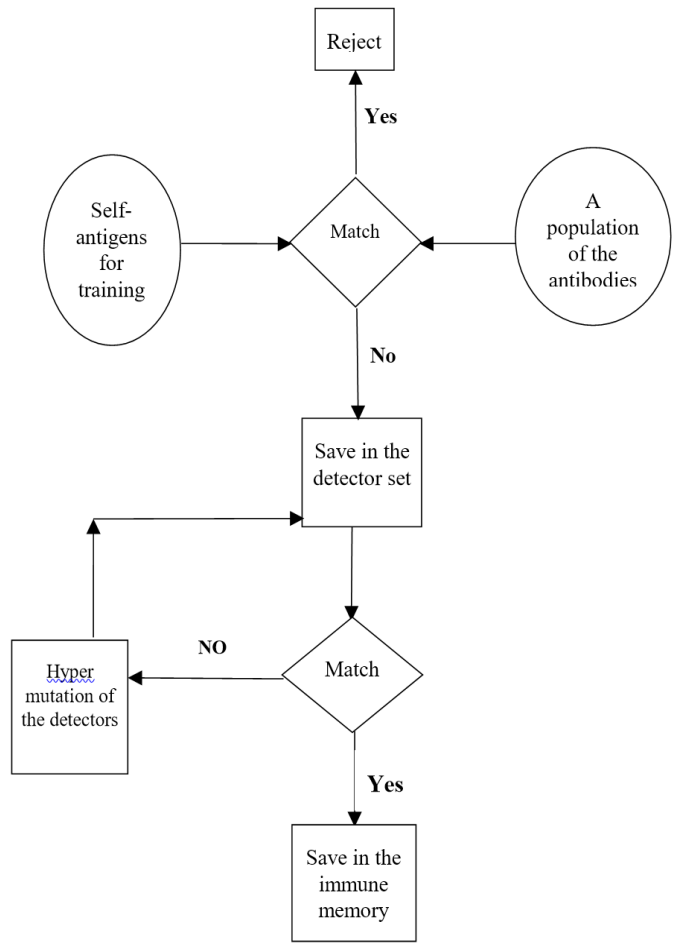

Fig. 8. Flowchart training of the detectors.

**Step 1: Initialization**

In this step, for each route that received RREP would be examined in terms of security. For this purpose, a test packet will be sent over each route and the destination node should send a confirmation packet. Obviously, if a route has a malicious node, the test packet would not reach to the destination and therefore confirmation packet would not be received. In this case, the probability of being a malicious node in the route will increased. The test packet will be sent three times to increase accuracy of the initialization step.

**Step 2: AIS-base detection**

As mentioned, the black hole attack shows hop count much lower than actual amount; so, in the case of having smaller hop count by a route, the probability of infection of this route would be increased. In similar manner, a route that has been infected by black hole attacker. Therefore, the desirable route is a route, which based on Equation (1), maximizes $F_r$ indicator.

$$Fr\ (HOP.Iteration) = \left(\frac{hop\ count}{Max\ hop\ count}\right) + \left(\frac{Max\ iteration\ node\ RREP}{Iteration\ node\ RREP}\right) \quad (1)$$

In order to combine results obtained from Step 1 and Step 2, considering metric security and efficiency, general an indicator of route selection obtains from following Equation (2).

Definition 1 ($P_{bh}(r)$) In this paper, a variable, termed as the route infection probability, is used to detect malicious nodes.

$$Seure\ Route = (1 - Pbh(r)) * Fr\ (HOP.Iteration) \quad (2)$$

Algorithm (1), shows training of detectors in which it is teaching a set of antibodies in order to be resistant with a set of self-antigens for the purpose of being supported to defend against non-self-antigen (black hole attack).

| **Algorithm 1:** For training the detectors (add in the step to the detector set) |
|---|
| 1: **Input**   : *S* = set of patterns to be recognized, <br> *n* the number of worst elements to select for removal |
| 2: **Output** : *M* = set of memory detectors capable of classifying unseen patterns |
| 3: **Begin** |
| 4:   Create an initial random set of antibodies, *A* |
| 5: **For All**  *patterns in S*   do |
| 6: Determine the affinity with each antibody in *A* |
| 7: Generate clones of a subset of the antibodies in *A with* the highest affinity. |
| 8: The number of clones for an antibody is proportional to its affinity |
| 9: Mutate attributes of these clones to the set *A* , and place a copy of the highest affinity antibodies in *A*   into the memory set, *M* |
| 11: Replace the *n* lowest affinity antibodies in *A* with new randomly generated antibodies |
| 12: **End** |
| 13: **End** |

According to Algorithm 2. For every route. Algorithm 2 are calculated and that route is selected as the most immune route, which has a bigger number meaning Hop and Iteretauion.

## 6. PERFORMANCE EVALUATION

In this part, we evaluate the performance of the proposed approach (AIS-DSR) in the problem of prevention black hole attack.

---

**Algorithm 2:** For selecting immune route

**Input:** $P_{bh}(r))$ for all routes

**Output:** An immune route

**Begin**

**Procedure** Selecting *immune route*

    R = number of routes

    For r = 1 To R

**If** Pm > 0.5 Then reject

**Else if** $F_r$ ( HOPCount, ITR) $=\frac{hop\ count}{Max\ hop\ count} + \frac{max\ Iteration\ nod\ rrep}{Iteration\ Node\ RREP}$

Select the route with maximum ROUTE i = (1 - Pm) * Fr (HOP COUNT ,ITR RREP)

**End if**

**End if**

**End procedure**

---

approach (AIS-DSR) in the problem of prevention black hole attack

### A. Performance metrics

We conduct extensive simulations to evaluate the effectiveness and performance of our AIS-DSR approach and compare it with DSR under black hole attack [32, 33 and 34]. We evaluate the throughput, end-to-end delay, packets loss ratio and packets drop ratio.

### 1. Throughput

Throughput is the number of data packets delivered from source to destination per unit of time. Throughput is calculated as received throughput in bit per second at the traffic destination. The throughput is calculated in Equation (1) follows: [35, 36 and 37].

$$\left(\frac{\sum_{j=1}^{n} Packets\ recieved}{\sum_{j=1}^{n} Packets\ originated}\right) \qquad (1)$$

### 2. End to end delay

This is the average delay between the sending of packets by the source and its receipt by the receiver. This includes all possible delays caused during data acquisition, route discovery, queuing, processing at intermediate nodes, retransmission delays at the MAC, propagation time, etc. It is measured in milliseconds. The lower value of end-to-end delay means the better performance of the protocol. The end-to-end delay is calculated in Equation (2) follows: [37, 38 and 39].

$$\left(\frac{\sum_{j=1}^{n} Delivery\ Time - \sum_{j=1}^{n} Arrival\ Time}{\sum_{j=1}^{n} Recieved\ packets}\right) \qquad (2)$$

### 3. Packet lost rate

Packet loss occurs when one or more packets of data traveling across a computer network fail to reach their destination. Packet loss is typically caused by network congestion. Packet loss is measured as a percentage of packets lost with respect to packets sent. The lower value of the packet loss means the better performance of the protocol. The PLR is calculated in Equation (3) follows:

$$PLR = \left(\frac{\sum_{j=1}^{n} Number\ of\ sent\ packets}{\sum_{j=1}^{n} Number\ of\ recieved\ packets}\right) * 100 \qquad (3)$$

### 4. Drop packet rate

Dropping packets, where malicious nodes drop packets and end up never forwarding them to a valid next hop. Thus, we can define DPR as shown in Equation (4) [39 and 40].

$$DPR = \left(\frac{\sum_{j=1}^{n} Number\ of\ droped\ packets}{\sum_{j=1}^{n} Number\ of\ droped\ packets + sent\ packets}\right) * 100 \qquad (4)$$

### B. Simulation results and analysis

We have implemented AIS-DSR approached in the NS-2 on Fedora 10. The simulation parameters are given in Table I.

Table 1. Simulation Parameters

| Parameters | Value |
| --- | --- |
| Channel Type | Channel/Wireless channel |
| Publication Type | Two ray ground |
| Network Interface | Wireless Phy |
| Antenna | Omni Antenna |
| Simulation Area (m x m) | 1000 X 1000 |
| MAC layer | MAC/802.11 |
| Traffic Type | CBR |
| Queue Type | Drop Tail |
| Number of nodes | 100 |
| Primary energy | 10 Jules |
| Threshold | 0.5 Jules |
| Time simulation | 200 s |

Fig. 9 compares the performance of AIS-DSR with that of DSR under black hole attacks for detection of the black hole attacks. As shown in the figure, AIS-DSR increases the throughput by more than 20% over than DSR under attacks, respectively.

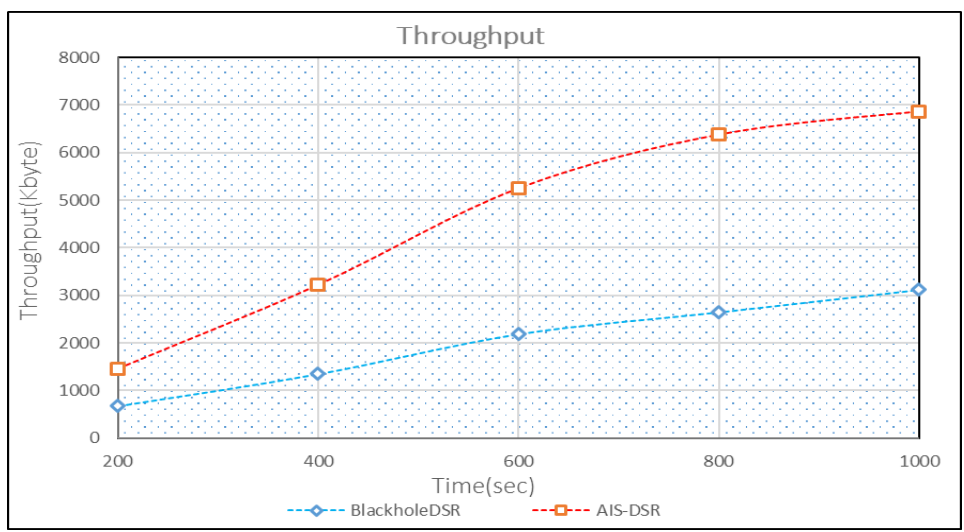

Fig. 9. Throughput vs. pause time

As in As shown in the Fig. 9, when there are destructive nodes in the network, the proposed algorithm by low end-to-end delay can identify the destructive nodes and aware another nodes, but DSR protocols are disable because it has more end-to-end delay.

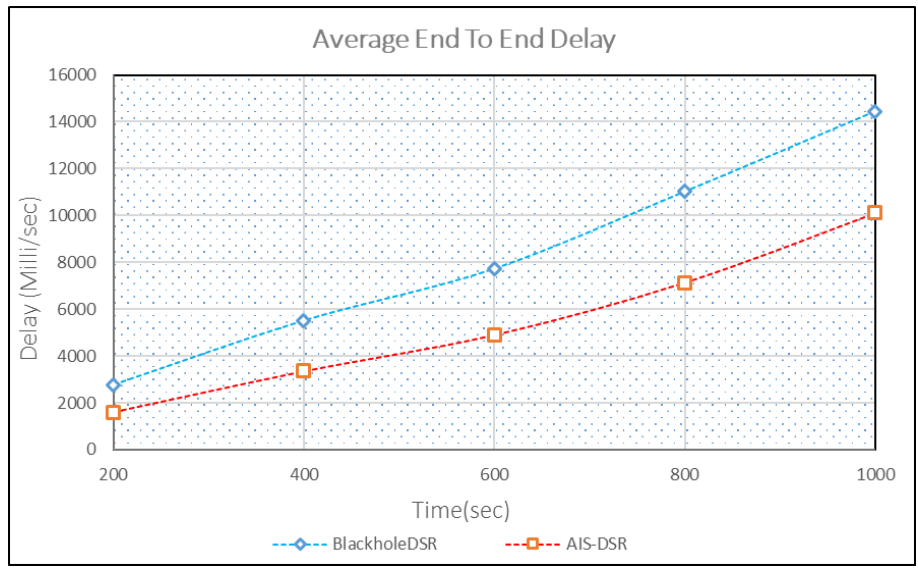

Fig. 10. End to end delay vs. pause time

DSR protocol has lost more packets than a AIS-DSR approach and it shows the success of destructive nodes in operating the attack of black hole on this protocol. In general, because of rules and computation of the proposed method for identifying the destructive nodes, the loss of packet rate in the proposed algorithm decreases compared to DSR protocol (Fig. 10).

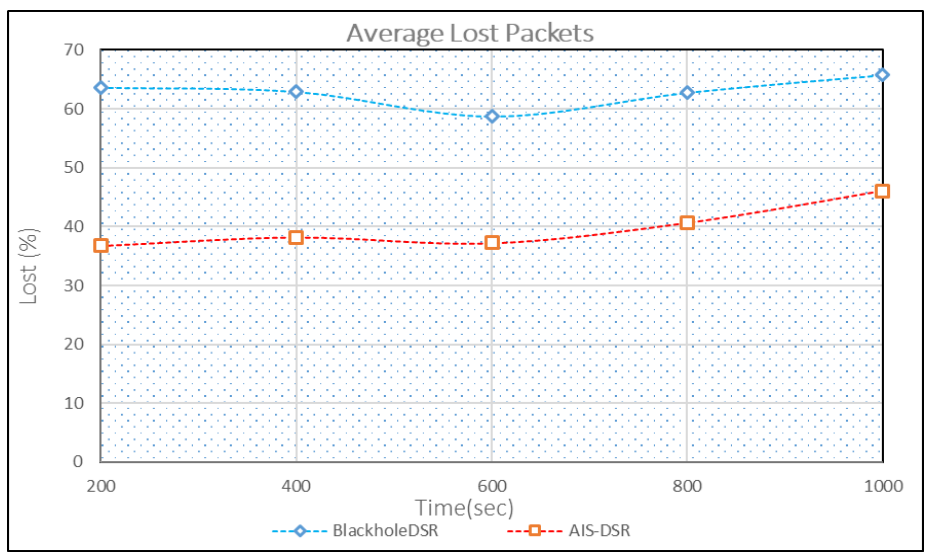

Fig. 11. Lost packets vs. pause time

Finally, Fig. 12 displays the dropped packets in the network. The proposed AIS-DSR technique drops a minimal number of packets compared to the other technique.

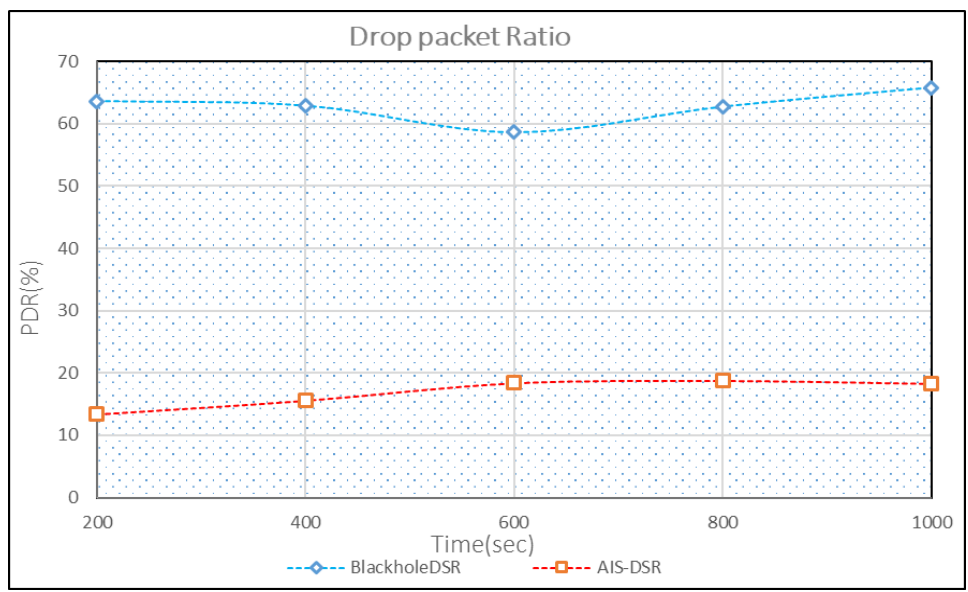

Fig. 12. Drop packets vs. pause time

## 7. Conclusion

In this paper, we proposed an approach for detecting black hole nodes in MANETs based on the artificial immune system. First, the artificial immune system was employed to develop a learning approach to detect and bypass the black hole attackers without affecting the overall performance of the MANETs. The aim of the proposed algorithm in this paper is that we can identify the isolate node and delete them from routing in accordance to the behavior of nodes in the system.

We compared the efficiency of our defensive scheme i.e. AIS-DSR with DSR under black hole attack. Simulation results showed that, in average, the overall performance of AIS-DSR is around 20% better than DSR routing protocol in terms of throughput, end to end delay, packet drop ratio and lost packets. In general, we can say that the proposed algorithm has better operation against malicious nodes than the DSR protocol.

In our future work, we plan to devise a mechanism to secure MANETs against the third attack model and test it intensively under different network conditions.

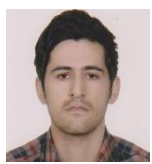
[41] Shahram Behzad received his B.Sc. degree in Computer Software Engineering from Islamic Azad University, Parsabad Branch and M.Sc. degree in Computer Software Engineering from Islamic Azad University, Shabestar Branch, under the supervision of Dr. Shahram Jamali Associate Professor University of Mohaghegh Ardabili, From 2013. From 2012 until now, he is a Lecturer in the Department of Computer Engineering, IAU University, Iran. His research interests include Computer Networks: Mobile and Wireless Ad-hoc Networks, Internet Protocols, Routing and MAC Layer Protocols, Artificial Immune System (AIS).

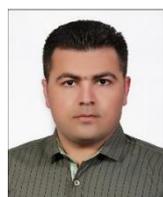
Reza Fotohi is currently a Ph.D. candidate in the Faculty of Computer Science and Engineering, Shahid Beheshti University, Tehran, IRAN. His research interests include Computer Networks, Mobile/Wireless Networks, UAV communications, IoT Security, Blockchain. He has over 8 years' research experience in computer networks and mobile/wireless networks. He was also the recipient and co-recipient of various awards for research publications. He is author and co-author of more than 20 technical journal and conference papers. He is also a member and reviewer in many journals and conferences such as IEEE Communications Surveys and Tutorials, IEEE Access, Applied Soft Computing,

Artificial Intelligence Review, Human-centric Computing and Information Sciences, The Journal of Supercomputing, Springer, Journal of Ambient Intelligence and Humanized Computing, Wireless Personal Communications Journal, National Academy Science Letters, Springer, Journal of Grid Computing, KSII Transactions on Internet and Information Systems, IET Signal Processing, International Journal of Communication Systems, Iranian Journal of Fuzzy Systems (IJFS), and The Turkish Journal of Electrical Engineering & Computer Sciences (ISI-JCR). His papers have more than 517 citations with 16 h-index and 21 i10-index.

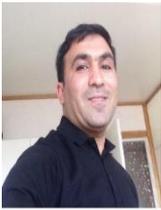 Jaber hosseini balov received his B.Sc. degree in Computer Software Engineering Technology from Urmia Academic Center for Education, Culture and Research (ACECR) and M.Sc. degree in Information technology and Management from Jönköping University in Sweden, under the supervision of Professor Anders Carstensen Associate Professor Computer Science in Jönköping University. From 2012 until now, he is a Lecturer in the Department of Computer Engineering, Urmia University of Applied Science and Technology in IRAN and Jönköping High school in Sweden.

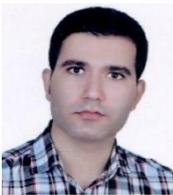 Mohammad javad Rabipour received his B.Sc. degree in Electrical Engineering, Communication from Najafabad Branch, Islamic Azad University, and M.Sc. degree in Electrical Engineering, Communication Systems from Kashan Branch, Islamic Azad University, under the supervision of Professor Hossein Ghasvari Associate Professor Electrical Engineering, Communication in Kashan Branch, Islamic Azad University, IRAN.